# An Open Community-Driven Model For Sustainable Research Software: Sustainable Research Software Institute

**Gregory R. Watson**[a], Addi Malviya-Thakur[a], Daniel S. Katz[b], Elaine M. Raybourn[c], Bill Hoffman[d], Dana Robinson[e], John Kellerman[f], Clark Roundy[f]

## Abstract

Research software plays a crucial role in advancing scientific knowledge, but ensuring its sustainability, maintainability, and long-term viability is an ongoing challenge. To address these concerns, the Sustainable Research Software Institute (SRSI) Model presents a comprehensive framework designed to promote sustainable practices in the research software community. This white paper provides an in-depth overview of the SRSI Model, outlining its objectives, services, funding mechanisms, collaborations, and the significant potential impact it could have on the research software community. It explores the wide range of services offered, diverse funding sources, extensive collaboration opportunities, and the transformative influence of the SRSI Model on the research software landscape.

---

[a] Oak Ridge National Laboratory, Oak Ridge, TN 37380
[b] University of Illinois, Urbana, IL 61801
[c] Sandia National Laboratories, Albuquerque, NM 87185
[d] Kitware Inc., Clifton Park, NY 12065
[e] The HDF Group, Champaign, IL 61820
[f] Eclipse.org Foundation Inc., Portland OR 97206

# 1. Introduction

Research software is essential for advancing scientific knowledge, but sustainability, maintenance, and long-term viability pose significant challenges. The Department of Energy (DOE) Exascale Computing Project (ECP)i[1] is a good example of this. The project has involved thousands of researchers who have developed millions of lines of open source code, but is slated to end in December 2023. As funding for development and maintenance of this software is now uncertain, DOE is examining options for sustaining this software for the foreseeable future. This white paper explores the Sustainable Research Software Institute (SRSI) Model, a comprehensive framework aimed at supporting sustainable research software development practices. Our goal is that this model, and its eventual implementation, will provide a means for DOE to sustain this valuable scientific software, and also provide an opportunity for other open source projects to participate in a vibrant, sustainable, research software ecosystem.

The SRSI Model envisions a sustainable, independent, and neutral organization dedicated to fostering excellence and innovation within the research software community, in part by sustaining important research software. In this model, SRSI comprises two main organizational components that collectively provide services and support to the research software ecosystem (see Figure 1). The first, the community services arm, described in Section 2, is focused on providing services to community programs which are a self-governing group of projects supported by contributors. The second, the sustainability funding arm, is focused on providing sustainability funding to support the ongoing sustainability activities of projects, and is described in Section 3. For various funding-related reasons that we won't go into in this paper, this is could be technically a separate legal organization, the SRSI Foundation.

---

[1] https://www.exascaleproject.org

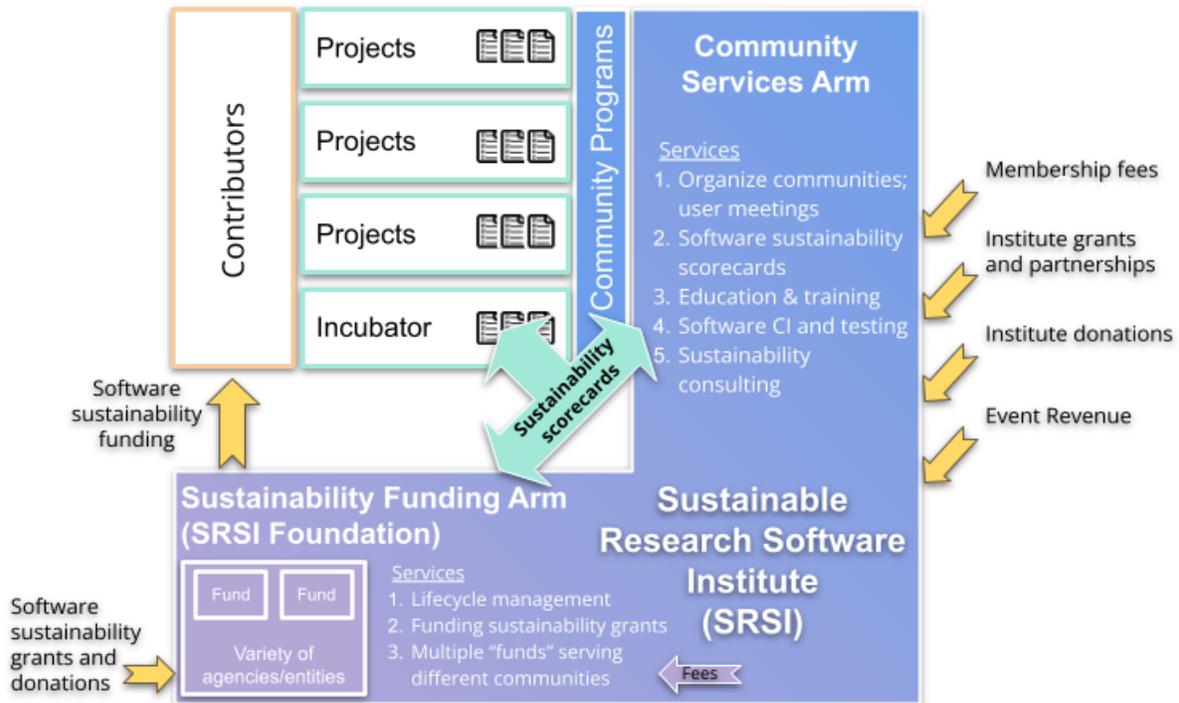

**Figure 1**: *Overview of the SRSI Model. The SRSI comprises two arms: community services (blue) and sustainability funding (purple) that collectively provide services and support to the research software ecosystem. The community services arm provides services to community programs which are a self-governing group of projects supported by contributors. The sustainability funding arm provides funds to support the ongoing sustainability activities of the projects. Funding is managed through a separate organization: the SRSI Foundation. Projects maintain sustainability scorecards which guide projects towards more sustainable practices while providing objective criteria for the SRSI Foundation to use for funding decisions.*

# 2. Community Services Arm

## 2.1 Objectives

The community services arm of the SRSI Model is primarily focussed on providing shared, community-wide, collective services that support the day-to-day activities of research software projects and their broader communities. The primary objectives of the community services arm include facilitating collaboration, organizing communities, providing education and training services, supporting software continuous integration (CI) and testing, and offering sustainability consulting services.

## 2.2 Community Programs

The goal of the SRSI Model is to create and foster an ecosystem in which research software projects can thrive and grow. SRSI recognizes the importance of communities in organizing and shaping the projects that are important to them, and so incorporates the notion of a *Community Program*. SRSI imposes minimal requirements on community programs and the research software projects within them: they must have a documented open, transparent, self-governance model for collaborative community operation. Embracing the principles of self-governance, the SRSI Model recognizes the value of community autonomy and decentralized decision-making processes. It encourages the development of an environment where community members feel a sense of ownership and responsibility, allowing them to collectively shape the direction of their research software projects. This involves setting up effective governance structures, like committees or working groups, that promote inclusivity and transparency when tackling important issues, defining project goals, and making informed decisions. By adopting a merit-based approach, the SRSI Model cultivates an environment that fosters excellence, innovation, and dedication, providing an equal opportunity for all community members to thrive. This ethos ensures fairness and inclusivity by rewarding and recognizing expertise and exceptional accomplishments within the research software project ecosystem.

### 2.2.1 Community Program Structure

SRSI aims to support community program structures that meet the needs of the communities. This means that communities can decide how best to represent their constituent projects. This may involve a deeper hierarchy (sub-communities) or other structure that helps the communities maintain coherence and ensure that everyone's voice is being heard.

### 2.2.2 Projects

Research software projects are at the heart of community programs. Projects can vary in size and scope depending on the needs of the community, and will typically have contributors from a variety of organizations. Projects must have a governance structure, either as defined by the community they belong to or by the project itself.

### 2.2.3 Incubator

Projects and communities join SRSI through an incubation process[2]. A new or existing community (and its associated projects) can join SRSI as an incubating community program. The community must have a documented governance structure and each project will need to undertake a sustainability assessment to create scorecards. Individual projects can also join as SRSI incubating projects. A community or project can graduate from incubation by achieving a basic level of sustainability and undertaking a graduation review. Incubating communities can join other community programs, or where appropriate, establish a new community program. A graduating project can join a community program that provides the best alignment for the project.

---

[2] During the startup phase of SRSI, a number of community programs and projects will be transitioned directly into the organization. See [Transitioning ECP Software Technology into a Foundation for Sustainable Research Software](#) for more details.

### 2.2.4 Community Assets

Assets, such as GitHub repositories, are the cornerstones of project activities. The SRSI Model ensures that these assets will be collectively owned and operated in an open and transparent manner for the life of the projects.

### 2.2.5 Intellectual Property

Projects maintain documented licensing and copyright policies that follow open-source software best practices. This ensures the ability to easily contribute to projects, and is an important facet of the SRSI Model.

## 2.3 Services

### 2.3.1 Organizing Communities and User Meetings

SRSI will bring together researchers, developers, and domain experts through organized communities and user meetings, promoting knowledge exchange, collaboration, and networking. These events serve as vital platforms for sustainable research software projects within specific research areas or disciplines.

### 2.3.2 Education and Training Services

The provision of a wide array of education and training programs to enhance the skills and knowledge of researchers and developers in sustainable software development practices will also be an important function of SRSI. These services will include workshops, courses, webinars, and resources covering topics such as software engineering best practices, version control, testing, and project management.

### 2.3.3 Software CI and Testing

The use of continuous integration (CI) and comprehensive testing practices are essential to ensuring the reliability and maintainability of research software, and form a fundamental part of the sustainability of software projects. SRSI will provide projects with access to CI infrastructure that will enable effective CI workflows, automated testing, and quality assurance processes that will empower researchers and developers to create robust and sustainable software.

### 2.3.4 Sustainability Consulting Services

In addition to the basic services provided to member projects, SRSI will also provide personalized consulting services to research software projects, offering guidance on various aspects of sustainability, including project management, software architecture, and long-term maintenance strategies on a cost-recovery basis. Through tailored consultations and workshops, these consultants will be available to assist projects in implementing sustainable development practices aligned with their specific needs.

## 2.4 Software Sustainability Scorecards

One of the primary activities of SRSI is to develop and implement software sustainability scorecards, providing comprehensive evaluations of research software projects based on objective, actionable, best-practice criteria such as code quality, documentation, community engagement, and long-term maintenance strategies. These scorecards offer valuable insights into project sustainability and maintainability, guiding researchers and developers in improving their software practices. The scorecards will be utilized by the SRSI Foundation to make decisions on where to target funding.

### 2.4.1 Scorecard Development Process

SRSI will be responsible for developing sustainability scorecards using the Scorecard Development Process (SDP). The SDP will utilize an open and collaborative process that leverages member and community knowledge of best practices, and build on other efforts to understand what makes a project sustainable. The SDP will result in the creation (and refinement) of a series of objective and actionable criteria that collectively provide a snapshot of a project's sustainability profile.

### 2.4.2 Scorecard Structure

Sustainability scorecards will be used to evaluate a project from a sustainability perspective using a series of objective best-practice criteria in a tiered badging system. Where possible, criteria that can be automatically determined will be used in order to simplify the evaluation process. Assessment criteria will include the importance/impact of the software, the project risk profile, factors relating to the project community, the maturity of the software engineering and project lifecycle practices, as well as other criteria gathered through community consultation.

### 2.4.3 Scorecard Use

Sustainability scorecards will have multiple beneficial uses, and projects will be strongly motivated to use them. At a minimum, scorecards will provide projects with insights into vulnerabilities in their sustainability and maintenance practices, and simplify how projects will be able to respond to these challenges. Additionally, scorecards will help project contributors learn about practices and techniques they can employ to the benefit of their project communities. Scorecards will also enable funding for sustainability practices to be directed to where it is most needed, and ensure that funding is distributed on an equitable basis and where it will produce the greatest impact.

## 2.5 Operational Funding

SRSI will sustain its activities through various funding sources, including membership fees, grants and partnerships with funding organizations, donations from individuals and corporations, and revenue generated from organizing events such as conferences, workshops, and training programs.

### 2.5.1 Membership Fees

It is envisioned that SRSI will operate under a tiered membership-based model, collecting membership fees to support its operational costs, services, and programs. Membership classes exist for entities wishing to be involved in SRSI corporate governance and policy development, as well as a free tier for entities who are primarily interested in project contributions.

### 2.5.2 Grants and Partnerships

SRSI will actively seek grants and partnerships with funding organizations, both governmental and non-governmental, that share its commitment to sustainable research software development. These grants and partnerships provide essential financial support for specific projects, initiatives, or programs within the SRSI, leveraging the expertise and resources of partnering organizations.

### 2.5.3 Donations

Individuals, corporations, and philanthropic foundations passionate about advancing research software sustainability can contribute to SRSI through donations. These generous contributions directly support SRSI's mission, enabling the expansion of its scope, enhanced services, and greater collaboration within the research software community.

### 2.5.4 Event Revenue

SRSI will organize revenue-generating events, such as conferences, workshops, and training programs. Registration fees, sponsorships, exhibition fees, and merchandise contribute to financing the institute's activities, while also providing opportunities for organizations and companies to engage with the research software community, supporting the sustainability efforts of SRSI.

# 3. Sustainable Funding Arm (SRSI Foundation)

## 3.1 Objectives

As the grant-making arm of the SRSI Model, the SRSI Foundation (SRSIF) focuses on funding research software sustainability activities. Its methods for doing this include funding sustainability grants, providing lifecycle management support, and facilitating collaborative funding partnerships.

## 3.2 Services

### 3.2.1 Funding Sustainability Grants

SRSIF offers financial support to the organizations contributing to research software projects and community programs through sustainability grants. These grants are awarded to projects that demonstrate a commitment to sustainable practices through the use of sustainability scorecards, and that have developed plans to enhance the long-term viability and maintainability of their software.

### 3.2.2 Lifecycle Management Support

SRSIF can provide lifecycle management support to community programs who desire this, assisting them in how to determine where to direct sustainability funds that would be most beneficial to the community, such as which research software projects to support for short-term or long-term grants. By utilizing software sustainability scorecards, input from community programs, and a streamlined proposal process, SRSIF can help make more informed decisions regarding where to allocate funds that would be most beneficial for sustainability of the software.

### 3.2.3 Collaborative Funding Partnerships

SRSIF actively encourages collaboration between funding agencies to establish joint funding programs. These partnerships enable funders to pool resources and expertise, fostering cross-disciplinary collaborations and expanding the reach and impact of their funding initiatives.

## 3.3 Funding Sources

SRSIF manages funds tailored to specific research areas, disciplines, or sustainability themes. These funds are used to provide financial support to the organizations contributing to the projects and communities in SRSI. Funding sources include grants from organizations, donations from individuals and institutions, fees charged for services, and collaborative funding partnerships.

### 3.3.1 Grants

SRSIF seeks grants from funding organizations that share its vision for sustainable research software development. Federal agencies or other funding bodies can provide grants to SRSIF that will be managed on their behalf and distributed to contributing organizations based on objective funding criteria. Individual communities and projects use SRSIF as a neutral party for submitting funding proposals and receiving and managing funds that will be directed to contributing organizations for that particular community/project. These grants provide primary financial resources for the foundation's initiatives, supporting research software projects and advancing software sustainability within the research community.

### 3.3.2 Donations

Donations from individuals, corporations, and institutions passionate about supporting sustainable research software, both specific projects and in general, are crucial for SRSIF's work. These contributions directly support the foundation's mission, providing financial assistance to projects, offering lifecycle management support, and fostering collaborations within the research software community.

### 3.3.3 Fees

SRSIF may charge fees for specific services, such as consulting or training, generating additional revenue to sustain its operations alongside other funding sources.

# 4. Implementation

This white paper describes a model for a community focused organization to promote and support sustainability of research software. We have identified three approaches for realizing this model in practice. These are:

1. *Creating a new organization*. This will require substantial investment in both startup activities, including establishing a legal entity, developing initial bylaws and community policies, hiring staff, building an online presence and necessary infrastructure, and other activities in order to be able to accept federal funds, and ongoing activities, including compliance, accounting, reporting, etc.

2. *Leveraging an existing foundation*. A number of open-source software foundations (e.g., NumFOCUS, etc.) already exist to foster a sustainable software ecosystem. One approach would be to join with one of these foundations to leverage the existing legal and organizational structure.

3. *Hosted by an existing foundation*. Some existing foundations (e.g., the Linux Foundation) provide a mechanism to host sub-foundations within their organizational structure. This differs from the previous two options because although it still requires creating a separate entity, the hosting organization provides much of the structure and expertise, along with a variety of resources to simplify and streamline the process.

We will not be examining the advantages and disadvantages of each approach here, but will be providing additional documents describing the possible implementations in the future. Due to the imminent end of ECP, we plan to focus on options 2 and 3 only.

# 5. Collaboration and Impact

The SRSI Model emphasizes collaboration and partnerships within the research software community. By joining SRSI and supporting its initiatives, both projects and funding agencies can benefit in several ways.

## 5.1 Benefits for Projects

### 5.1.1 Funding Opportunities

Joining SRSI makes projects eligible for funding opportunities, and by participating in SRSI's funding mechanisms, projects can access financial support to further their research software development, implement sustainability improvements, and ensure long-term viability. Projects are still free to seek funding from other sources for sustainability or non-sustainability related activities.

### 5.1.2 Access to Comprehensive Support

Under the SRSI umbrella, projects access to a wide range of services and resources tailored to support sustainable research software development. These include opportunities for knowledge exchange, collaboration, and networking through organized communities and user meetings. Projects can connect with researchers, developers, and domain experts in their research areas, fostering collaboration and improving the quality of their software.

### 5.1.3 Improved Software Sustainability

SRSI's software sustainability scorecards provide valuable insights into project sustainability and maintainability. By evaluating code quality, documentation, community engagement, and long-term maintenance strategies, projects can identify areas for improvement and implement best practices, leading to more sustainable software. Projects also benefit by learning about sustainability practices they may not have been aware of.

### 5.1.4 Enhanced Skills and Knowledge

Education and training services offered by the SRSI equip researchers and developers with skills and knowledge in sustainable software development practices. Through workshops, courses, webinars, and resources, projects can improve their understanding of software engineering best practices, version control, testing, and project management, enabling them to build more robust and maintainable software.

### 5.1.5 Guidance and Consultation

The sustainability consulting services provided by SRSI offer personalized guidance on various aspects of sustainability. Projects can receive tailored advice on project management, software architecture, and long-term maintenance strategies, aligning their development practices with sustainability principles for more viable and impactful software.

## 5.2 Benefits For Funding Entities

### 5.2.1 More Sustainable Software Development

By supporting SRSI's initiatives, funding agencies demonstrate their commitment to sustainable research software development, aligning with objectives related to scientific knowledge advancement and innovation.

### 5.2.2 Increased Impact of Funding

Collaborating with SRSI allows funding agencies to pool resources and expertise through joint funding programs. This collaborative approach fosters cross-disciplinary collaborations and expands the reach and effectiveness of funding initiatives, leading to significant advances in research software development.

### 5.2.3 Streamlined Project Selection

SRSIF's lifecycle management support can help funders' financial contributions target research software projects that will benefit most from their support. By utilizing software sustainability scorecards, community input, and a streamlined proposal process, funders can make more informed decisions, ensuring funded projects align with sustainability principles and have long-term viability and impact.

### 5.2.4 Diverse Funding Opportunities

SRSIF manages funds tailored to specific research areas, disciplines, or sustainability themes. This diversity allows funders to support projects closely aligned with their interests and objectives, expanding their reach within the research community.

### 5.2.5 Collaboration and Knowledge Sharing

SRSI fosters collaboration and knowledge sharing between funding agencies, research institutions, and the research software community. This collaborative environment enables funders to learn from each other, share best practices, and collectively advance sustainable research software development. The SRSI serves as a platform for exchanging ideas and promoting collective learning, enhancing the effectiveness and impact of funding efforts.

# 6. Conclusion

This paper provides a high level overview of the Sustainable Research Software Institute (SRSI) Model for realizing a sustainable approach to developing research software. Under this model, SRSI comprises a community services arm and a sustainable funding arm (SRSIF), which together provide a comprehensive framework to support sustainable research software development. By offering a wide range of services, engaging diverse funding sources, promoting collaboration, and leveraging community input, the SRSI Model strengthens the research software ecosystem. SRSI benefits both projects and funders, leading to improved software sustainability, enhanced skills and knowledge, access to funding opportunities, and increased collaboration and impact. Active participation in the SRSI Model contributes to scientific knowledge and innovation through sustainable research software development practices.

## Acknowledgement

This manuscript has been authored by UT-Battelle, LLC under Contract No. DEAC05-00OR22725 with the U.S. Department of Energy. The United States Government retains and the publisher, by accepting the article for publication, acknowledges that the United States Government retains a nonexclusive, paid-up, irrevocable, world-wide license to publish or reproduce the published form of this manuscript, or allow others to do so, for United States Government purposes. The Department of Energy will provide public access to these results of federally sponsored research in accordance with the DOE Public Access Plan (http://energy.gov/downloads/ doe-public-access-plan).